\documentclass[usenatbib]{mn2e}

\usepackage[dvips]{graphicx}
\usepackage{amssymb}
\usepackage{txfonts}

\newcommand{\msun}{{\rm M}_{\sun}}
\newcommand{\rsun}{{\rm R}_{\sun}}

\newbox\grsign \setbox\grsign=\hbox{$>$} \newdimen\grdimen \grdimen=\ht\grsign
\newbox\simlessbox \newbox\simgreatbox \newbox\simpropbox
\setbox\simgreatbox=\hbox{\raise.5ex\hbox{$>$}\llap
     {\lower.5ex\hbox{$\sim$}}}\ht1=\grdimen\dp1=0pt
\setbox\simlessbox=\hbox{\raise.5ex\hbox{$<$}\llap
     {\lower.5ex\hbox{$\sim$}}}\ht2=\grdimen\dp2=0pt
\setbox\simpropbox=\hbox{\raise.5ex\hbox{$\propto$}\llap
     {\lower.5ex\hbox{$\sim$}}}\ht2=\grdimen\dp2=0pt
\def\ga{\mathrel{\copy\simgreatbox}}
\def\la{\mathrel{\copy\simlessbox}}
\def\simprop{\mathrel{\copy\simpropbox}}

\title[Orbital modulation of radio emission in Cyg X-1]{The structure of the jet in Cyg X-1 inferred from orbital modulation of the radio emission}

\author[A. A. Zdziarski]
{Andrzej A. Zdziarski\\
Centrum Astronomiczne im.\ M. Kopernika, Bartycka 18, PL-00-716 Warszawa, Poland\\
}

\date{Accepted 2012 February 13.  Received 2012 February 13; in original form 2011 May 21}

\pagerange{\pageref{firstpage}--\pageref{lastpage}}
\pubyear{2012}

\begin{document}

\maketitle

\label{firstpage}

\begin{abstract}
We study free-free absorption of radio emission by winds of massive stars. We derive formulae for the optical depth through the wind measured from a point of emission along a jet, taking into account Compton and photoionization heating and Compton, recombination, line and advection cooling. 

We apply the developed formalism to radio monitoring data for Cyg X-1, which allows us to obtain strong constraints on the structure of its inner jet. With the data at 15 GHz, and taking into account an anisotropy of the stellar wind in Cyg X-1, we estimate the location of the peak of that emission along the jet at about one orbital separation, i.e., $\sim 3\times 10^{12}$ cm. Given a previous determination of the turnover frequency in Cyg X-1, this implies the location of the base of the jet at $\sim 10^3$ gravitational radii. We also obtain corresponding results at 8.3 GHz and 2.25 GHz, which roughly follow the standard conical partially self-absorbed jet model. Furthermore, we find that the level of the orbital modulation depends on the radio flux, with the modulation being substantially stronger when the radio flux is lower. This is explained by the height of the radio emission along the jet decreasing with the decreasing radio flux, as predicted by jet models. Based on the finding of the flux-dependent orbital modulation, we are able to estimate a range of the possible changes of the form of the radio/X-ray correlation in Cyg X-1 due to free-free absorption. We also derive predictions for the orbital modulation and flux attenuation at frequencies beyond the 2.25--15 GHz range. 
\end{abstract}

\section{Introduction}
\label{intro}

Cyg X-1 is an archetypical and well studied persistent black-hole binary. Its companion is the OB supergiant HDE 226868. Cyg X-1 is both a radio and X-ray source, and the fluxes in these two bands are strongly correlated (e.g., \citealt*{gfp03,z11}, hereafter Paper I). The likely cause of the correlation is the jet in this system (\citealt{stirling01}, hereafter S01) being formed by the matter of an inner hot accretion flow \citep*{hs03,mhd03,yc05}. The radio emission is modulated at the orbital period \citep*{pfb99}, which is caused by free-free absorption in the stellar wind from the massive companion. As shown by \citet{sz07} (hereafter SZ07), the observed modulation levels imply that the average radio fluxes are substantially absorbed, which then, as discussed in Paper I, modifies the form of the observed radio/X-ray correlation. 

In this work, we study in detail the free-free absorption of radio emission in winds of massive stars in high-mass X-ray binaries (hereafter HMXBs). We derive formulae for the optical depth measured from the point of emission along the jet. We also consider emission of counter-jets and Compton scattering by the wind.

Then, we apply our results to Cyg X-1, taking into account the strong anisotropy of its stellar wind, found by \citet{gies08}, as well as the radio emission being extended (S01). This yields strong constraints on the structure of the jet in Cyg X-1, in particular it implies that a major part of this emission takes place at distances comparable to the binary orbit. Based on the results of this application, we estimate the effect of the free-free absorption on the form of the radio/X-ray correlation in Cyg X-1. 

\section{Radiative heating and free-free absorption in a stellar wind}
\label{wind}

A  standard wind velocity profile from a massive star is (e.g., \citealt*{lcp87}),
\begin{equation}
v(r) \simeq v_{\infty}\left(1-{R_* \over r}\right)^{\beta},
\label{velocity}
\end{equation}
where $r$ is the distance from the centre of the donor of radius $R_*$, $v_\infty$ is the terminal velocity, and $\beta$ parameterizes the wind acceleration. Here, a small correction due to velocity reaching the sound speed rather than being null at the stellar surface has been neglected. The ion density is, 
\begin{equation}
n(r) = {-f(\vartheta,\varphi) \dot M \over 4 \upi m_{\rm p}\mu_{\rm i} r^2 v(r) },
\label{n_r}
\end{equation}
where $f$ is the direction-dependent factor accounting for the wind asymmetry due to its focusing (see below), $\dot M$ is the total mass-loss rate, $\vartheta$ and $\varphi$ (measured from the axis connecting the stars) are the polar and azimuthal angle, respectively, $m_{\rm p}$ is the proton mass, $\mu_{\rm i}\simeq 4/(1+3X)$ is the mean ion molecular weight, and $X$ is the H fraction. For simplicity, we hereafter assume $X=1$.

The local wind temperature follows from the balance of radiative heating and cooling and advection. In an optically-thin approximation \citep{km82}, applicable here, the radiative rates depend only on the shape of the irradiating spectrum and on the ionization parameter,
\begin{equation}
\xi \equiv {L_{\rm ion}\over n x^2},
\label{xi}
\end{equation}
\citep*{tts69}, where $x$ is the distance to the irradiating source with the ionizing luminosity, $L_{\rm ion}$. The net rate of X-ray photoionization heating and recombination cooling per ion has been fitted as \citep{blondin94},
\begin{equation}
C_{\rm X}\simeq 1.5\times 10^{-21}\xi^{1/4} T^{-1/2}(1-T/T_{\rm X})n
\label{C_X}
\end{equation}
(in units of erg s$^{-1}$), where $T$ is in K and $kT_{\rm X}\simeq 10$ keV is the characteristic energy of the ionizing spectrum. Given that the wind temperature in Cyg X-1 has $kT\ll 1$ keV, we can neglect the second term in parentheses above. The formulae of \citet{blondin94} have been derived for the case of a 10-keV bremsstrahlung X-ray spectrum in a high-mass X-ray binary. The rate of bremsstrahlung and line cooling has been fitted as \citep{blondin94},
\begin{eqnarray}
\lefteqn{-C_{\rm b,l}\simeq 3.3\times 10^{-27} T^{1/2} n(r)+\nonumber}\\
\lefteqn{\qquad\quad
\left[1.7\times 10^{-18}\exp(-T_{\rm l}/T)\xi^{-1}T^{-1/2}+10^{-24}\right]n(r)}
\label{C_bl}
\end{eqnarray}
where $T_{\rm l}=1.3\times 10^5$ K. 

\begin{figure}
\centerline{\includegraphics[width=6.5cm]{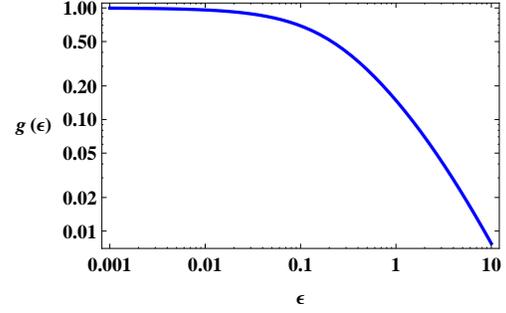}}
\caption{The function $g(E/m_{\rm e} c^2)$, by which the Compton heating of electrons with $kT\ll m_{\rm e} c^2$ by photons with energy $E$ is reduced due to the Klein-Nishina effects. 
}
\label{f:g_e}
\end{figure}

In the non-relativistic regime, the rate of Compton energy exchange between electrons and photons is proportional to $\int (E-4 kT)F(E){\rm d}E$. The rate per electron including Klein-Nishina corrections but assuming $kT\ll m_{\rm e} c^2$ can be written as,
\begin{equation}
C_{\rm C}=
{\sigma_{\rm T}\over m_{\rm e} c^2} \int g(\epsilon) E F(E) {\rm d}E - {4 kT\sigma_{\rm T}\over m_{\rm e} c^2} \int h(\epsilon) F(E) {\rm d} E,
\label{Compton}
\end{equation}
where $F(E)$ is the differential energy flux, $\sigma_{\rm T}$ is the Thomson cross section, $m_{\rm e}$ is the electron mass, $\epsilon\equiv E/m_{\rm e} c^2$, and $g(\epsilon)$ and $h(\epsilon)$ give the Klein-Nishina corrections to Compton heating and cooling, respectively. The former is given by \citep{guilbert86,bi95}, 
\begin{eqnarray}
\lefteqn{g(\epsilon)= \frac{3 (\epsilon^2-2\epsilon-3) \log (1+2\epsilon)}{8 \epsilon^4}+\frac{9+51\epsilon+93\epsilon^2+51\epsilon^3-10\epsilon^4}{4 \epsilon^3 (1+2\epsilon)^3} \nonumber}\\
\lefteqn{\qquad \simeq 1-{21 \epsilon\over 5}+{147 \epsilon^2\over 10}+O(\epsilon^3). \label{g_e}}
\end{eqnarray}
Separating $F(E)$ into the stellar and X-ray component, we have,
\begin{equation}
C_{\rm C}=  {\sigma_{\rm T}\over \upi m_{\rm e} c^2}\left[
{1\over 4 x^2 } \int\! g(\epsilon) E L_{\rm X}(E) {\rm d}E+{\eta kT_* L_* - kT (L_*+L_{\rm X}) \over  r^2}\right],
\label{Compton2} 
\end{equation}
where $L_*$ is the stellar luminosity, $T_*$ is the stellar temperature, $L_{\rm X}(E)$ is the differential luminosity originating at the black-hole location, $r$ and $x$ is the distance to the centre of the donor and to the compact object, respectively, $\eta=90 \zeta(5)/\upi^4\simeq 0.96$, and $\zeta$ is the Riemann zeta function. The X-ray heating term in equation (\ref{Compton2}) is valid at any values of $\epsilon$, provided $kT/m_{\rm e}c^2\ll 1$. Equation (\ref{Compton2}) also assumes that cooling is dominated by photons at $\epsilon\ll 1$, which, e.g., is the case when $L_*\gg L_{\rm X}$. If it is not satisfied, $h(\epsilon)$ needs to be included as well, see eq.\ (17) in \citet{bi95}. The function $g(\epsilon)$ is plotted in Fig.\ \ref{f:g_e}. We see it decreases rather fast at $E\ga 50$ keV. At a point $(r,\vartheta,\varphi)$, 
\begin{equation}
x^2=r^2+a^2-2 a r \sin\vartheta\cos\varphi,
\end{equation}
where $a$ is the separation between the stars. 

Then, assuming the velocity profile of equation (\ref{velocity}), we need to solve the energy equation, which, for an ionized gas, can be written as,
\begin{equation}
v(r)\left[3k{{\rm d}T\over {\rm d}r}+2 k T n(r) {{\rm d}\over {\rm d}r}{1\over n(r)}\right] =C_{\rm X}+C_{\rm b,l}+C_{\rm C},
\label{energy}
\end{equation}
This can be solved for $T(r)$ adopting a boundary condition close to the stellar surface, where the wind is optically thick and our assumption of optical thinness breaks down. In an optically thick wind, \citet{cb06} find $T\sim 0.6T_*$. We have found that the solution above the assumed boundary, $r>r_0$, is virtually independent of the value of $r_0$, and we adopt $r_0=(1.2$--$1.3) R_*$. At low radii, ${\rm d}T/{\rm d}r>0$, and cooler particles are brought to a given point from radii closer to the star. At all radii, adiabatic expansion cools the wind. Thus, the dominant effect of the left-hand side of equation (\ref{energy}) is cooling. We have found that solutions of this equation are rather well approximated by replacing the differential terms (i.e., its left-hand side) by an algebraic advection cooling term, $-C_{\rm adv}$, see, e.g., \citet*{cal97} for a similar approach. We have found that we achieve the highest accuracy in this approach defining $C_{\rm adv}$ as,
\begin{equation}
-C_{\rm adv}={3 kT v_\infty\over r}. 
\label{advection}
\end{equation}
We thus solve an algebraic equation for $T(r)$,
\begin{equation}
C_{\rm X}+C_{\rm b,l}+C_{\rm C}+C_{\rm adv}=0.
\label{T_eq}
\end{equation}
See Section \ref{cygx1} for an example comparison of the two solutions.

\begin{figure}
\centerline{\includegraphics[width=6.3cm]{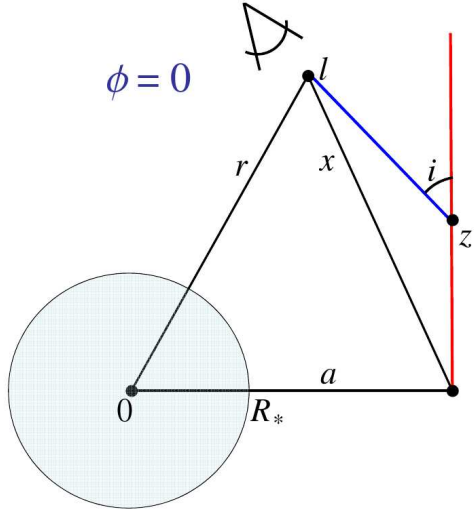}}
\caption{A schematic representation of the considered system at the superior conjunction. The distance between the high-mass donor (which centre is at 0) and the compact object is $a$. The point of radio emission is at the height $z$ along the jet (shown by the red line), which originates at the compact-object location. A point along the photon path through the wind towards the observed is at the distance, $l$, from the emission point, at the distance, $r$, from the centre of the donor, and at the distance, $x$, from the compact object, which is surrounded by the source of X-rays.
}
\label{f:geo}
\end{figure}

We then assume the bulk of the radio emission at a given $\nu$ occurs at a distance, $z$, along a jet from its origin at the compact object, see Fig.\ \ref{f:geo}. This $z$ corresponds to the place along the jet where it becomes optically thin to synchrotron self-absorption (\citealt{bk79}, hereafter BK79). The jet is assumed to be perpendicular to the circular orbital plane, see Fig.\ \ref{f:geo}. We note that the assumption of perpendicularity is not a crucial; the jet can be inclined (as suggested for Cyg X-1 by \citealt*{mbf09}) or curved (as in the Cyg X-1 model of SZ07), which introduces a phase lag to an orbital modulation. Then, $z/a$ is to be interpreted as the height above the orbital plane in units of the distance to the rotation axis of the donor. In this case, $a$ would be somewhat higher than the orbital separation. We neglect here this complication. We also neglect any effects of the jet opening angle, as a second-order effect. In the case of Cyg X-1, it has been constrained to be $\la 2\degr$ (S01), which is negligibly small for our calculations.

As shown in Fig.\ \ref{f:geo}, $l$ is the path length from the emission location along the jet to a point at a radius, $r$, from the donor centre. We find, 
\begin{equation}
(r/a)^2=1+(z/a)^2+(l/a)^2- 2 (l/a) \left[\sin i\cos\phi- (z/a)\cos i\right], \label{r2}
\end{equation}
where $i$ is the binary inclination and $\phi$ is the orbital phase, which equals $\phi=0$, $\upi$ at the superior and inferior conjunction, respectively. Then,
\begin{equation}
(x/a)^2=(z/a)^2+(l/a)^2+ 2(z/a) (l/a) \cos i. 
\label{x2}
\end{equation}

The free-free absorption coefficient  at a frequency, $\nu$, is,
\begin{equation}
\alpha_{\rm ff} \simeq 0.12 T^{-3/2}  
n^2 \nu^{-2}\, {\rm cm}^{-1}\, {\rm cm}^{-1},
\label{alpha_ff}
\end{equation}
where $T$ is in units of K and $\nu$ is in GHz. At a reference temperature, $T=T_0$, we can write $\alpha_{\rm ff}=\alpha_0 (n_{\rm i}/n_0)^2$, where $\alpha_0$ is the absorption coefficient at an ion density of $n_0$. We define $n_0$ as the density at $r=a$ under the assumption of $v=v_\infty$. We also define a normalization optical depth of $\tau_0=\alpha_0 a$.

Then, the free-free optical depth is,
\begin{equation}
\tau=\tau_0 \int_0^\infty \left(r\over a\right)^{-4} \left(1-{R_* \over r}\right)^{-2\beta} \left[T(r)\over T_0\right]^{-3/2}
{\rm d}(l/a).
\label{tau_v}
\end{equation}
This can be calculated numerically with $T(r)$ as the solution of either equation (\ref{energy}) or (\ref{T_eq}). 

On the other hand, we also consider the case of an isothermal wind with constant velocity (which is approached at at $r\gg R_*$). In that case, 
\begin{equation}
\tau\simeq \tau_0 \int_0^\infty \left(r\over a\right)^{-4} {\rm d}(l/a),
\label{tau_const}
\end{equation}
In this approximation, we obtain an analytical formula for $\tau$, 
\begin{eqnarray}
\lefteqn{ {\tau\over \tau_0} = {\upi t/2+ s u^{1/2}+  t \arctan(s u^{-1/2})\over 2 t u^{3/2} }
={t\, {\rm arccot}(s u^{-1/2})-s u^{1/2}\over 2 t u^{3/2}}, \nonumber}
\\
\lefteqn{t\equiv 1+(z/a)^2,\quad s\equiv \sin i\cos\phi- (z/a)\cos i,\quad u\equiv t-s^2. \label{tau_ff}}
\end{eqnarray}
We find $u>0$ everywhere except for $\phi=\upi$, $z/a=\cot i$, when $u=0$ and equation (\ref{tau_ff}) appears to have a singularity. However, $\tau$ is still finite there, $\tau/ \tau_0 = \sin^3 i/ 3$. For $z\rightarrow 0$,
\begin{eqnarray}
\lefteqn {{\tau\over \tau_0} = {\upi+ 2\mu+\sin 2\mu\over 4\cos^3\mu}+ O(z,i,\phi),\quad
\mu\equiv \arcsin(\sin i\cos \phi),}\\
\lefteqn{{\Delta\tau\over \tau_0}\equiv {\tau(\phi=0)-\tau(\phi=\upi)\over \tau_0} ={2i+\sin 2i\over 2 \cos^3 i}+ O(z,i),}
\label{z0}
\end{eqnarray}
where $\Delta\tau$ is the difference in the optical depths at the superior and inferior conjunctions. For large $z$, the leading order of $\tau$ is independent of $\phi$,
\begin{equation}
{\tau\over \tau_0} = {2i - \sin 2 i\over 4(z/a)^3 \sin^3 i}+O[(z/a)^{-4},i,\phi], \quad z/a\gg 1,
\label{tau_asymp}
\end{equation}
and $\Delta\tau$ is lower by one order of $a/z$,
\begin{equation}
{\Delta\tau\over \tau_0}={9\sin i-12 i\cos i+ \sin 3 i\over 4(z/a)^4 \sin^4 i}+O[(z/a)^{-5},i],\quad z/a\gg 1.
\label{dtau_asymp}
\end{equation}

Potential constraints from eclipses of the radio emission from both the jet and counter-jet are discussed in Appendix \ref{eclipses}. Appendix \ref{scattering} presents results for Compton scattering (or another process with the cross section proportional to the density) analogous to the above ones for free-free absorption.

We define the full modulation depth due to absorption as,
\begin{equation}
D \equiv {F(\phi_{\rm max}) - F(\phi_{\rm min}) \over F(\phi_{\rm max})}=1-\exp(-\Delta \tau), 
\label{ModDepth}
\end{equation}
where $F(\phi_{\rm max})$ and $F(\phi_{\rm min})$ are the maximum and minimum fluxes in the model averaged and folded light curves, observed at $\phi_{\rm max}$ and $\phi_{\rm min}$, respectively, and $\phi_{\rm min}$ represents the possible phase lag. In our model, $\phi_{\rm max}-\phi_{\rm min}=\upi$. We note that $D$ (or, equivalently, $\Delta \tau$) in the model is uniquely determined by $z/a$, $i$ and $\tau_0$. The attenuation averaged over the orbit and the average $\tau$ are defined as,
\begin{equation}
A= {\int_0^\upi \exp(-\tau){\rm d}\phi\over \upi}
\equiv \exp(-\bar{\tau}),
\label{av_tau}
\end{equation}
where the integration can be done over only 0--$\upi$ due to the planar symmetry. As a rough approximation for $\bar{\tau}$, $\tau(\upi/2)$ may be used. We note that $D$, $A$, $\bar{\tau}$, $\phi_{\rm min}$ are functions of frequency. 

\section{Parameters of the Cyg X-1 binary and its stellar wind}
\label{pars}

The ephemeris of Cyg X-1 has been determined, e.g., by \citet{lasala98}, \citet{brocksopp99}. It can be written as,
\begin{equation}
t_{\rm sup} [{\rm MJD}]=50077.995 + P E,
\end{equation}
where $P=5.599829$ d, $t_{\rm sup}$ is the time of a superior conjunction (the black hole furthest from the observer), and $E$ is an integer. 

Observational determinations of the mass of the donor, $M_*$, and of the black hole, $M_{\rm X}$, $R_*$, $i$, and the distance, $d$, are all mutually connected. We use here the constraints of \citet{cn09}, and of the evolutionary models of \citet{zi05}. \citet{cn09} obtained the most likely ranges of $M_*\simeq (17$--$31)\msun$, $M_{\rm X}\simeq (8$--$16)\msun$, $R_*\simeq (15$--$21)\rsun$, $i\simeq 43\degr$--$31\degr$, $d\simeq (1.5$--2.0) kpc. They also measured the donor temperature as $T_*\simeq 2.8\times 10^4$ K, and derived,
\begin{equation}
R_*/\rsun = (10.3\pm 0.7)d,\qquad i=\arcsin[1.02\pm 0.09)/d],
\label{r1_i}
\end{equation}
where $d$ is in kpc. Here, we choose their solution that also agrees with the detailed evolutionary modelling of \citet{zi05}, namely, $i=31\degr$, $d=2.0$ kpc, $M_*=30\msun$, $M_{\rm X}=16\msun$, $R_*=20.7\rsun$, which yields $a\simeq 47.5\rsun\simeq 2.3 R_*$\footnote{After the calculations reported where completed, \citet{reid11} have determined the trigonometric parallax distance as $d=1.86^{+0.12}_{-0.11}$ kpc (with $1\sigma$ uncertainties), which value is compatible with adopted here within $\sim 1\sigma$. Then, \citet{orosz11} determined $M_{\rm X}=14.8\pm 1.0\msun$, again within $1\sigma$ from our value. They also found $M_*=19.2\pm 1.9\msun$, $R_*=16.2\pm 0.7\rsun$, and $i=27.1\pm 0.8\degr$. The effect of these difference on our results have been found to be relatively minor, and less important than the uncertainty regarding the structure of the stellar wind.} The eccentricity is $\ll 1$ \citep{orosz11}, and we here assume the orbit to be circular.

The total mass loss rate in the hard spectral state was found by \citet{gies03} as $-\dot M\simeq 2.6\times 10^{-6}\msun$ yr$^{-1}$. However, the mass loss can be highly asymmetric in binary systems, especially those in which the donor is close to filling its Roche lobe. This is the case in Cyg X-1, where $R_* >0.9$ of the Roche-lobe radius \citep{gb86a,gb86b,gies03,zi05}. In this situation, the wind is focused towards the black hole \citep{fc82}, with the wind density inside the Roche lobe being substantially higher than that outside it. The presence of the focused wind component was found to fit very well the He {\sc ii} $\lambda$4686 emission profile of Cyg X-1 \citep{gb86b}. Evidence for the focused component based on soft X-ray lines was also found by \citet{miller05} and \citet{hanke09}.

Further complexity of the wind geometry was found by \citet{gies08}. Namely, they found the unfocused, locally spherically symmetric wind component emitted by the X-ray irradiated hemisphere of the donor to be undetectable. This weakness is apparently due to wind ionization by X-rays, which strongly reduces the wind radiative acceleration by the lines emitted by the donor. On the other hand, the other hemisphere of the donor, shaded from the X-rays, emits a normal wind, similar to that of an isolated massive star \citep{blondin94}. Summarizing, the wind of HDE 226868 has three main components: the wind focused along the axis joining the stars, which carries most of the mass loss; the standard radiatively driven wind from the hemisphere in the shadow, and a weak wind component from the donor hemisphere facing the black hole. As found by \citet{gies08}, this third component is weak in both spectral states, hard and soft, of Cyg X-1.

Radio photons are emitted by the jet relatively high above the orbital plane and may be attenuated mainly by this third weak component, i.e., the quasi-spherical wind from the hemisphere facing the X-ray source. Given its weakness, it has $f\ll 1$, but no specific constraints are currently available. For the wind velocity profile, we use here that of equation (\ref{velocity}) with $v_\infty\simeq 1.6\times 10^8$ cm s$^{-1}$ and $\beta= 1$ \citep{gb86b}. 

The stellar luminosity for our assumed parameters is $L_*\simeq 9.1\times 10^{38}$ erg s$^{-1}$. We then use the average hard X-ray/soft $\gamma$-ray spectrum of Cyg X-1 as observed by the OSSE and BATSE detectors on board of {\it Compton Gamma Ray Observatory\/} \citep{mcconnell02}. We fit this spectrum with an e-folded power law, $F_{\rm X}(E)=K (E/1\,{\rm keV})^{-\alpha} \exp(-E/E_{\rm f})$, obtaining $K=0.62$ cm s$^{-1}$ (normalized to the OSSE flux), $\alpha=0.36\pm 0.03$, $E_{\rm f}=160\pm 5$ keV, yielding $L_{\rm X}\simeq 1.7\times 10^{37}$ erg s$^{-1}$. The fit is based on the data at $\geq 20$ keV, and it quite severely underestimates the actual spectrum in softer X-rays. However, the main contribution to Compton heating is from hard X-rays, see equation (\ref{Compton2}), and thus this underestimate has a negligible effect. We note that the Klein-Nishina correction is very important for this spectrum, with the Compton heating rate calculated non-relativistically, with $g(\epsilon)\equiv 1$, being 2.7 times higher than the correct rate calculated using $g(\epsilon)$ of equation (\ref{g_e}). On the other hand, photoionization is mostly due to soft X-rays, which spectrum in Cyg X-1 is relatively uncertain due to absorption. We assume here $L_{\rm ion}=1\times 10^{37}$ erg s$^{-1}$ in equation (\ref{xi}). 

Then, we can express the characteristic free-free optical depth of the wind, $\tau_0$ in equation (\ref{tau_v}), as,
\begin{eqnarray}
\lefteqn{
\tau_0\simeq 35.5 \left( - f \dot{M}\over 2.6\times 10^{-6}\msun\,{\rm yr}^{-1}\right)^2 \left(T_0\over 10^6\,{\rm K}\right)^{-3/2} \left(M_*+M_{\rm X}\over 46\msun\right)^{-1} \nonumber}\\
\lefteqn{\qquad\times  \left(v_\infty\over 1.6\times 10^8\,{\rm cm\ s}^{-1}\right)^{-2} \left(\nu\over 15\,{\rm GHz}\right)^{-2},  
\label{tauref}}
\end{eqnarray}
where we used $10^6$ K as the reference temperature. Given the lack of a measurement of the wind in the polar regions \citep{gies08}, we treat $f$ as a free parameter of the model. 

\section{Radio emission in the hard state of Cyg X-1}
\label{radio}

We study here radio monitoring data at 15 GHz from the Ryle Telescope, which cover MJD 50226--53902, and the Arcminute Microkelvin Imager (AMI), MJD 54573--55540. Also, we use the 2.25 GHz and 8.3 GHz monitoring data from the Green Bank Interferometer, for MJD 50409--51823. These data are the same as in Paper I, and we refer to it for details. We fit only data corresponding to the hard spectral state, which corresponds to MJD 50350--50590, 50660--50995, 51025--51400, 51640--51840, 51960--52100, 52565--52770, 52880--52975, 53115--53174, 53540--53800, 53880--55375 (which intervals contain 12408 Ryle/AMI  measurements and 1160 GBI ones), see \citet*{pzi08} and Paper I. 

An important issue here is that these telescopes do not spatially resolve the radio emission. On the other hand, the emission in the hard state is known to be extended. S01 have found that $\sim 1/3$--1/2 of emission at 8.4 GHz observed by VLBA is within the beam of a longitudinal extent of $\sim 3''$, whereas the remainder forms an extended jet visible up to $\sim 15''$. This has been confirmed by further VLBA observations at 8.4 GHz reported in \citet{rushton11}, in which the resolved fraction was $\sim 0.3$--0.4. The resolution of $3''$ corresponds to the length of $9\times 10^{13}(d/2,{\rm kpc})/\sin i$ cm, which is $\sim 30 a/\sin i$. The resolved 8 GHz emission can thus have only a tiny orbital modulation. 

The observational situation at 15 GHz is more complex. The VLBA beam size of the observations reported in \citet{rushton09} is $\sim 1.3''$, which corresponds to $\simeq 4\times 10^{13}(d/2,{\rm kpc})/\sin i$ cm $\simeq 10a/\sin i$. The unresolved and total fluxes for the three observations taken on the same days as the 8.4 GHz ones of S01 (i.e., MJD 51035, 51037, 51039) are (7.3, 7.7) mJy, (10.9, 12.2) mJy, (6.1, 7.8) mJy. Thus, the unresolved fractions are relatively small, 0.05--0.22. However, Cyg X-1 was also monitored by the Ryle Telescope three times on each of those days, yielding the total fluxes of (13.2, 14.9, 16.0 mJy), (13.2, 12.2, 10.4 mJy), (18.9, 19.4, 18.1 mJy), respectively, i.e., substantially more than the total VLBA fluxes for the 1st and 3rd observation. Although all of the Ryle observations were taken $\sim 0.1$ d before the start of the VLBA observations, Cyg X-1 was in a relatively stable state at that time with no large-amplitude variability detected. It thus appears that the VLBA measurement missed an extended part of the resolved emission. Then, the resolved fractions are $\simeq 0.1$--0.7. 

As discussed in \citet*{zls12}, it is unlikely that the radio jet in Cyg X-1 follows a single set of power-law dependencies from its base to the resolved parts. It appears that it contains a compact part within $\sim (1$--$10) a$, and then a secondary dissipation event leads to formation of a more extended part of the jet, resolved by VLBA. The observed strong orbital modulation (see below) has to occur in the compact part. The division of the fluxes between the two parts as a function of time remains unknown. Based on the above data, we can only roughly estimate that about a half of the flux is emitted by the compact part. Given that the radio-IR spectrum of Cyg X-1 appears to be a single power law \citep{fender00,rahoui11}, both components have approximately the same spectra. 

\section{Free-free absorption in Cyg X-1}
\label{cygx1}

With the model developed in Section \ref{wind}, we can fit the orbital modulation profiles, which are given by the light curves averaged and folded over the orbital period. In this way, we obtain the location of the bulk of the emission, $z/a$. We neglect the emission of the counter-jet, as its emission is beamed away from the observer and it passes through a much higher column density than that of the jet, including the dense focused component close to the binary plane. Given the results discussed in Section \ref{radio}, we split the radio light curves into unmodulated and modulated parts,
\begin{equation}
F(t)= b F(t)_{\rm unmodulated}+(1-b)F(t)_{\rm modulated},
\label{split}
\end{equation}
where $t$ it time. The second term on the right-hand side corresponds to emission of the inner part of the jet, within $\sim 10^1 a$ or so, which is orbitally modulated. The first term corresponds to the larger-scale emission, which is not modulated. Then,
\begin{equation}
{\bar F}_{\rm intr}(t)={\bar F}(t)\left[b+(1-b) \exp(\bar{\tau})\right],
\label{fintr}
\end{equation}
where ${\bar F}_{\rm intr}$, ${\bar F}$ denote the light curves averaged over time intervals of the length $P$. Given that the observed emission comes from a region of the size $\la 10^{15}$ cm (S01; \citealt{rushton09,rushton11}) and a relativistic speed of the jet (S01; \citealt{mbf09}), we assume that both parts of $F_{\rm intr}$ vary in the same way (apart of the orbital modulation) on a $\ga 10^{4.5}$-s time scale, below which scale the radio variability is weak.

Equation (\ref{fintr}) implies the average attenuation of
\begin{equation}
A={1\over b+(1-b) \exp(-\bar{\tau})}.
\label{av_taub}
\end{equation}
The modulation depth becomes,
\begin{equation}
D \equiv {1-\exp(-\Delta\tau) \over 1+b\exp(\tau_{\rm min}-\bar\tau)/(1-b)},
\label{ModDepthb}
\end{equation}
where $\tau_{\rm min}$ is the minimum optical depth through the wind (corresponding to the maximum flux, at $\phi_{\rm max}$). 

In the calculations below, we assume either $b=0.5$ (see Section \ref{radio}) or $b=0$, which provide us with an estimate of the sensitivity of the calculations to taking into account the unmodulated part of the flux. The assumed $b=0.5$ is, given the available data, only approximate, as well as it is unlikely that it was constant during the course of the monitoring. In Figs.\ \ref{f:orb_mod}--\ref{f:orb_mod15_var} below, we show the results for $b=0$, which directly correspond to the observations of the modulation. After obtaining folded and averaged light curves of $F(t)$, we subtract from them $b\langle F\rangle$, as implied by equation (\ref{split}), and then recalculate the folded average modulation. 

We need to properly average the folded light curve. We are interested in variability governed by the orbital time scale, and not in that on much shorter time scales. Therefore, we rebin the observed light curve with a bin size equal $\Delta P=P/K$, where $K$ is the number of bins per period. In this way, we also partly remove a bias on the folded averages due to non-uniform coverage, e.g., due to a large numbers of points in some orbital bins and low number in other bins. The errors on the resulting averages are then calculated from dispersion of fluxes averaged within individual bins of the same phase range and the size $\Delta P$. This method was applied to periodic variability of Cyg X-1 by \citet*{izp07}. We fit $\int_{\Delta\phi} \exp(-\tau){\rm d}\phi/\Delta\phi$, where $\Delta\phi=2\upi/K$ is the phase bin width, instead of using the value at a bin centre. Also, finite phase lags with respect to the superior conjunction are observed in Cyg X-1. Then, the phase lag is a free parameter of the model, and $\tau(\phi)$ of equations (\ref{tau_v}--\ref{tau_ff}) is replaced by $\tau(\phi- \phi_{\rm min})$.

We first compare our results with those of SZ07 (whose results are for $b=0$). In that work, the atomic heating and cooling was treated more accurately than in this work. On the other hand, adiabatic cooling was not included, which resulted in a constant temperature, $T_0$, at large radii. Thus, we first compare their results with those using our analytical, constant $T$, model, equation (\ref{tau_ff}). Using the parameters from SZ07, namely, $f=1$, $T=3.2\times 10^5$ K and $i=40\degr$, we obtain $z/a\simeq 3.3$, fitted to the 15 GHz data, whereas SZ07 found $z/a\simeq 3.7$; thus there is a good agreement. On the other hand, we find $z/a\simeq 5.8$ using the full model of equation (\ref{tau_v}). The higher value found by us is due to our inclusion of adiabatic cooling, which reduces the temperature at large radii and thus increases the wind opacity there, requiring in turn the emission place at a higher height. 

However, although SZ07 used the same total $\dot M$ as that used here, they assumed isotropy, without taking into account focusing and the reduction of the wind from the irradiated surface of the donor. This resulted in a relatively high density of the wind. As a result, the 15 GHz emission of Cyg X-1 was found to be attenuated to $A\simeq 0.2$, see fig.\ 13 in SZ07. As also shown there, the average attenuation decreased with decreasing frequency at $\nu\la 30$ GHz, to $A\simeq 0.5$ at 2.25 GHz, as well as also decreased with increasing frequency at $\nu \ga 30$ GHz. However, as argued in Paper I, this is not compatible with the 2--220 GHz spectrum being rather straight with $\alpha\simeq 0$ \citep{fender00}, and with no substantial dip at 15 GHz, see also figs.\ 10--11 in Paper I. This supports the wind density in the polar regions being substantially lower than that for an isotropic wind, see Section \ref{pars}. Thus, the free-free optical depth of the wind is most likely substantially lower than that modelled by SZ07. Furthermore, the model of SZ07 implies a very strong orbital modulation, by $\sim$90 per cent in the 150--220 GHz range. This range was observed by \citet{fender00}, and no apparent orbital modulation was seen. In particular, there was one 146 GHz observation at $\phi/2\upi=0.95$, at which the measured flux was about equal to the average one. 

\begin{figure}
\centerline{\includegraphics[width=7.4cm]{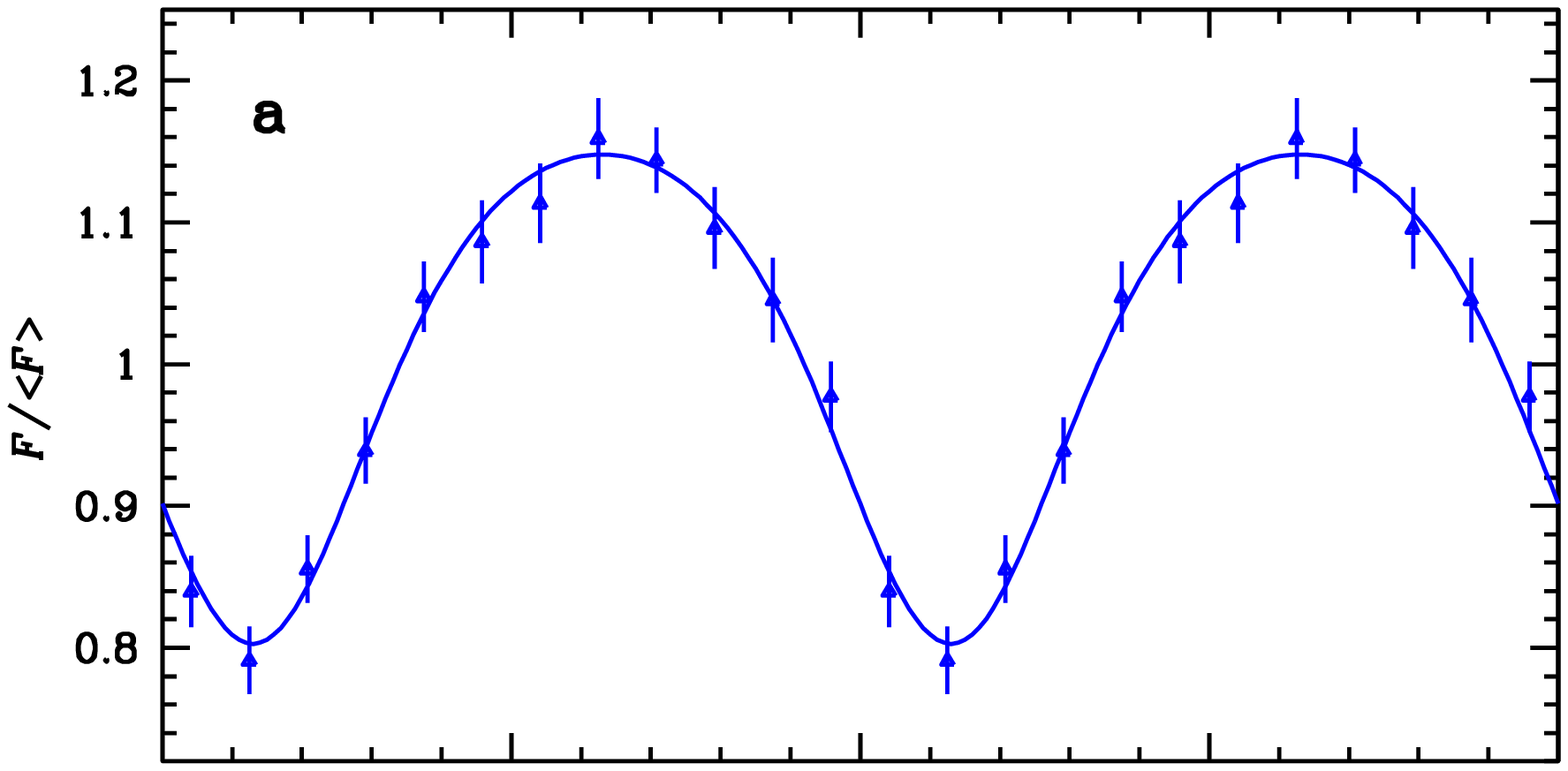}}
\centerline{\includegraphics[width=7.4cm]{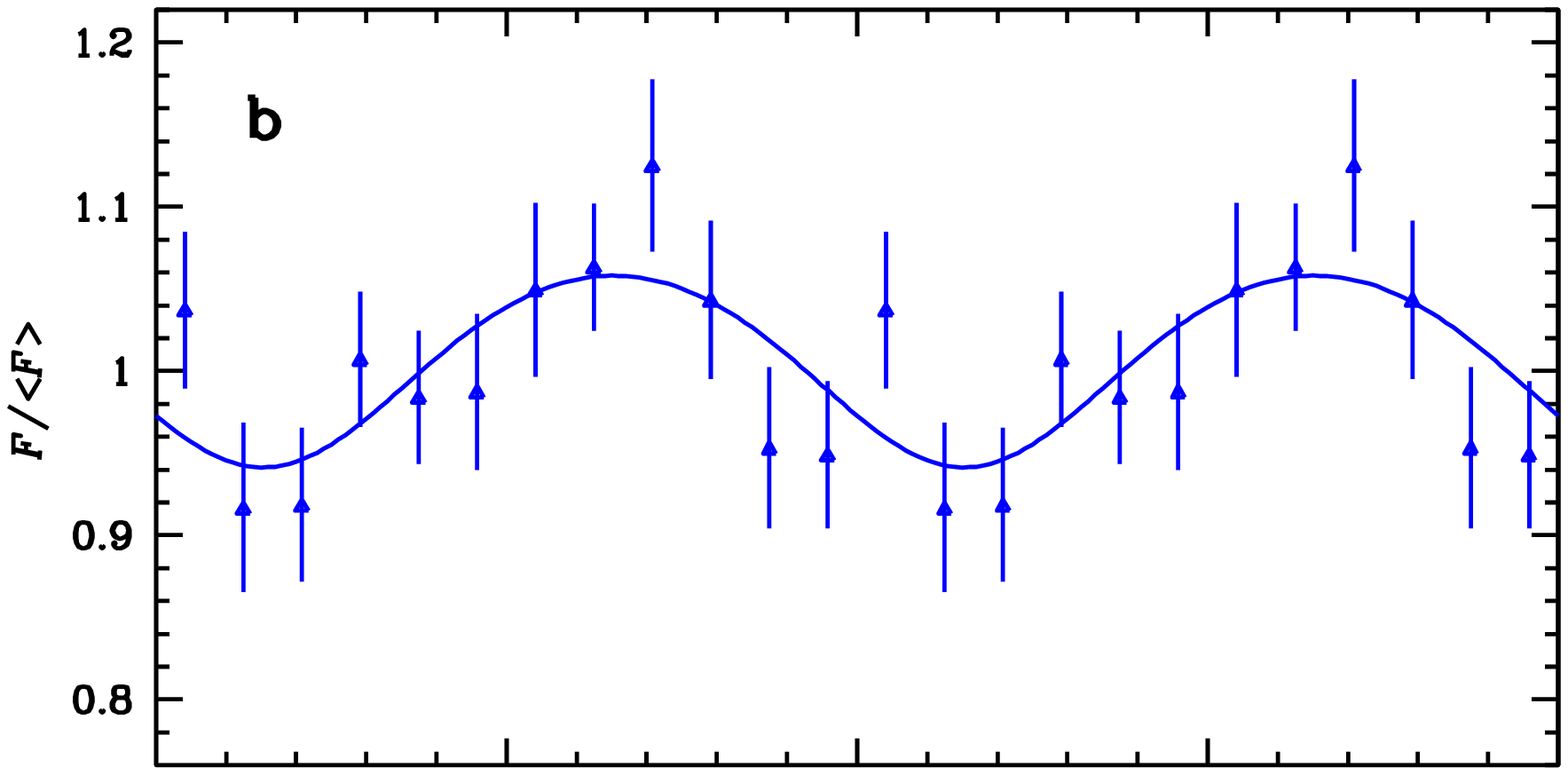}}
\centerline{\includegraphics[width=7.4cm]{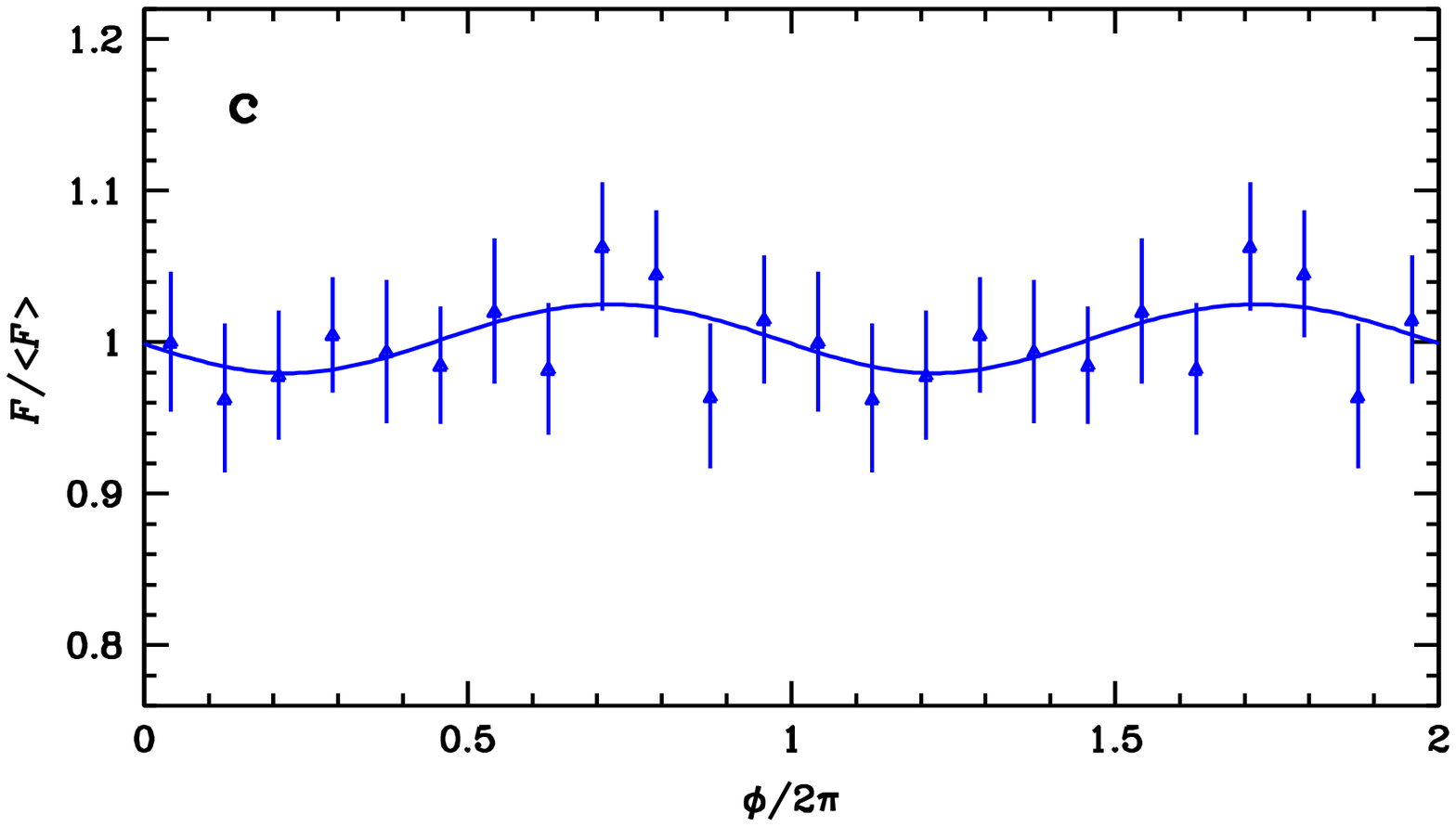}}
\caption{The observed (i.e., at $b=0$) orbital modulation (data points with errros) at (a) 15 GHz, (b) 8.3 GHz, (c) 2.25 GHz fitted by the model (solid curve) of  equation (\ref{tau_v}) with $i=31\degr$. The flux averages are $\langle F\rangle\simeq 11.6$ mJy, 14.2 mJy, 14.5 mJy, respectively. 
}
\label{f:orb_mod}
\end{figure}

\begin{figure}
\centerline{\includegraphics[width=7.4cm]{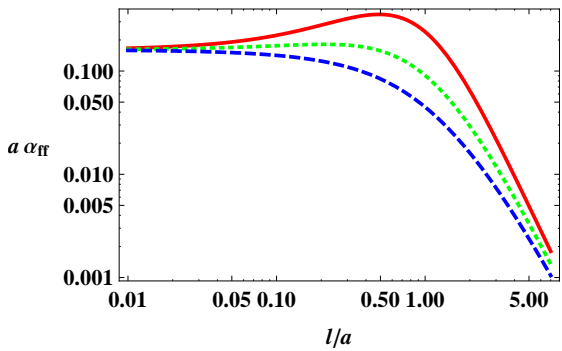}}
\caption{Variations of the free-free opacity along a photon path for our model for 15 GHz (at $b=0$, $f=0.06$, $i=31\degr$). The solid, dashed and dotted curves correspond to the orbital phase of $(\phi-\phi_{\rm min})/2\upi=0$, 0.5 and 1, respectively. 
}
\label{f:dtau}
\end{figure}

\begin{figure}
\centerline{\includegraphics[width=7.4cm]{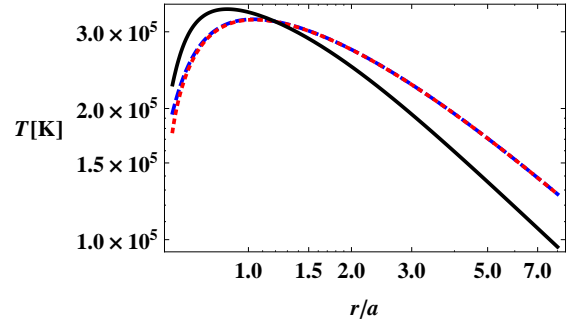}}
\caption{An example of the radial temperature profile for our model for 15 GHz (at $b=0$, $f=0.06$, $i=31\degr$) at the direction of $\vartheta=45\degr$, $\varphi=90\degr$. The dashed and dotted curves correspond to the solution of the differential energy equation (\ref{energy}) with the boundary condition imposed at $1.2R_*$ and $1.3R_*$, respectively, and the solid curve corresponds to the algebraic solution of equation (\ref{T_eq}).
}
\label{f:radial}
\end{figure}

\begin{figure}
\centerline{\includegraphics[width=7.4cm]{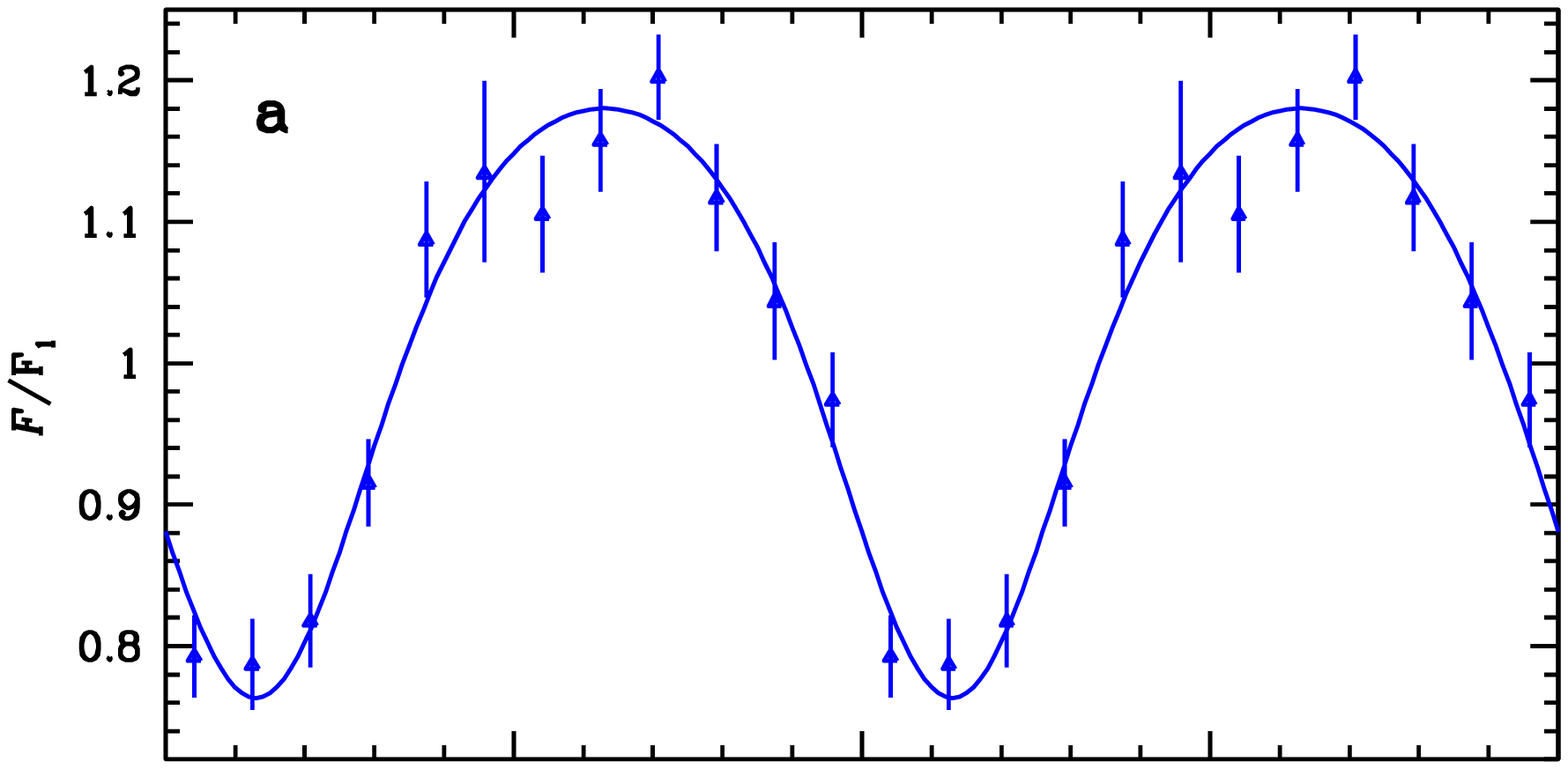}}
\centerline{\includegraphics[width=7.4cm]{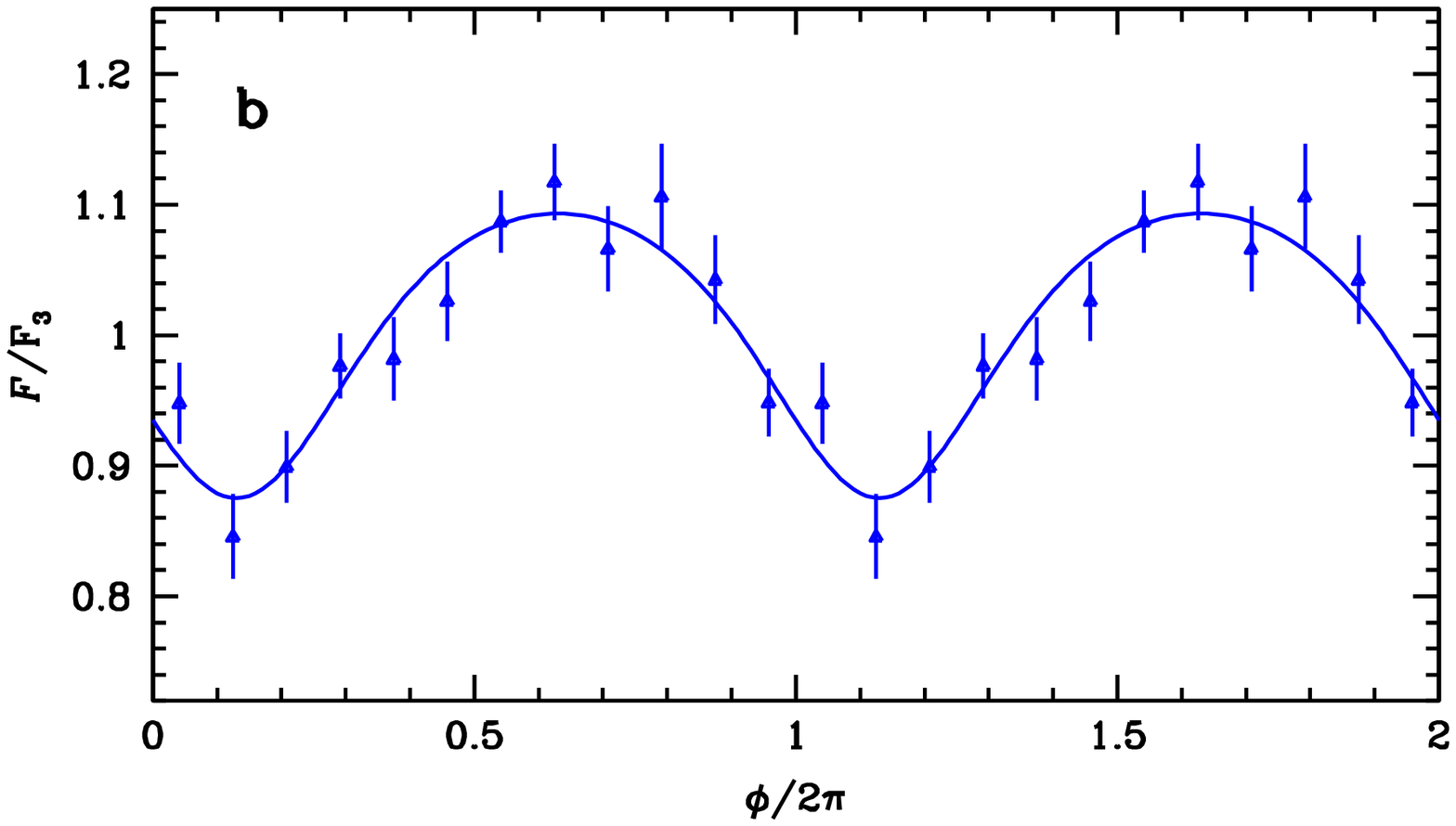}}
\caption{The observed ($b=0$) orbital modulation at 15 GHz fitted by the model of  equation (\ref{tau_v}) with $i=31\degr$ for the fluxes (a) $<\langle F\rangle/1.15$ and (b) $>1.15\langle F\rangle$, where $\langle F\rangle=11.6$ mJy is the average for the entire sample. 
}
\label{f:orb_mod15_var}
\end{figure}

\begin{figure}
\centerline{\includegraphics[width=7cm]{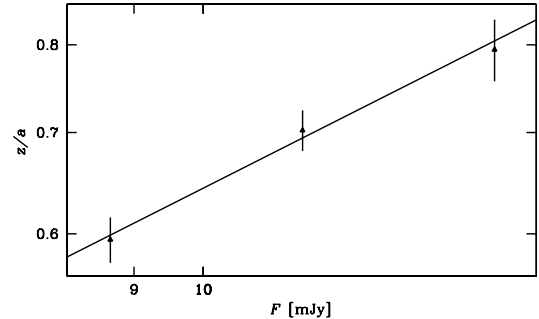}}
\caption{The dependence of the fitted height along the jet, $z$ (in units of $a$) on the 15 GHz flux, for the case $b=0.5$, $f=0.1$. The line shows the best fit.
}
\label{f:z_F}
\end{figure}

Our best data are those at 15 GHz, and we fit them with the model of equation (\ref{tau_v}). We first give the results neglecting the extended character of the radio emission, i.e., for $b=0$. We obtain $f\simeq 0.060\pm 0.005$, $z/a\simeq 0.57\pm 0.01$, $\phi_{\rm min}/2\upi\simeq 0.13\pm 0.01$, at $\chi^2_\nu\simeq 3.4/8$. Fig.\ \ref{f:orb_mod}(a) presents a fit of this model to the folded and averaged light curve. The model modulation depth is $D\simeq 0.30\pm 0.01$, and the orbit-averaged attenuation is $A=0.76\pm 0.01$. The values of $\phi_{\rm min}/2\upi$ and $D$ are very similar to those obtained by \citet{l06}, who used a smaller data set and a different model. 

Fig.\ \ref{f:dtau} shows variation of the absorption coefficient along the photon path, for our best-fit model. We see the opacity drops fast at $l\ga a$. The region $l\la a$ is at radii of $r\sim 1.5 a$. Fig.\ \ref{f:radial} shows an example of the radial temperature profile for the best-fit model. We see that indeed the solution is virtually independent of the adopted inner boundary condition and the differences between the solution of the differential energy equation and the approximate one with the advection term are relatively small, $\la 20$ per cent. 

We then consider the effect of taking into account the effect of a part of the radio flux being emitted well beyond the orbit, assuming $b=0.5$. We obtain the same values for $f$ and $\phi_{\rm min}$ as for $b=0$, and $z/a\simeq 0.23\pm 0.02$ at $\chi^2_\nu\simeq 6.0/8$, and with $D\simeq 0.31\pm 0.01$, $A=0.82\pm 0.01$. We note, however, that the obtained $z$ is relatively close to the binary plane, which may no more be in the polar region, and thus not to be consistent with the fitted low value of wind-density scaling factor, $f$. Thus, we also consider models with fixed larger values of $f$, but still $\ll 1$. At $f=0.1$, $z/a\simeq 0.62\pm 0.02$, $D\simeq 0.30\pm 0.01$, $A=0.70\pm 0.01$, $\chi^2_\nu\simeq 13.7/9$, and at $f=0.2$, $z/a\simeq 1.21\pm 0.02$, $D\simeq 0.29\pm 0.01$, $A=0.52\pm 0.01$, $\chi^2_\nu\simeq 27.5/9$. Since $z\sim a$ in both cases, the emission region is likely to be in the low-density polar region. Although these values of $\chi^2_\nu$ are rather high, the bad fit quality may be due to our lack of knowledge of the actual structure of the stellar wind and of the spatial distribution of the radio emission along the jet. Hereafter, we adopt $f=0.1$, where both $f\ll 1$ and $z\sim a$, but also use $f=0.2$ for comparison in some cases. 

Based on general features of the jet emission, e.g., BK79, \citet{hs03}, we expect that the height at which the bulk of emission at a given frequency emerges increases with increasing radio flux. This is due to the associated increase of the density of the relativistic electrons in the jet, which increases the synchrotron self-absorption optical depth. Assuming the magnetic field strength to be proportional to the equipartition value with the relativistic electrons, we derive (using the formalism of \citealt{zls12}), 
\begin{equation}
z\propto F_{\rm intr}^q,\quad q={s+6\over 2 s+13},
\label{zF}
\end{equation}
where $s$ is the index of the electron distribution ($\propto \gamma^{-s}$, where $\gamma$ is the Lorentz factor), which gives $q\simeq 0.47$ for $0.7< s<3.5$. [Equation (5) of \citet{heinz06} gives this result for $s=2$.] The increased height of the point where the bulk of emission occurs results in less wind absorption and a lower modulation depth. Also, a higher X-ray flux (which is positively correlated with the radio flux, e.g., Paper I) leads to a higher temperature, equation (\ref{T_eq}), and thus less wind absorption of radio photons. 

Given the good statistical quality of the 15 GHz light curve, we can test the above theoretical prediction with the data. For this purpose, we have divided the observed range of the 15 GHz flux in the hard state, $F$, into three parts, satisfying (1) $\bar{F} <\langle F\rangle/1.15$, (3) $\bar{F}>1.15 \langle F\rangle$, and (2) that in between. As the criterion, we have used the radio fluxes averaged over a single orbital period, $\bar{F}$, instead of individual measurements, $F$, since the latter are themselves affected by the orbital modulation. The unweighted geometric average fluxes in these ranges, $j=1$, 2 and 3 are $F_j=8.7$ mJy, 11.6 mJy and 15.6 mJy, respectively. In order to account for a part of the radio flux emitted at large distances, equation (\ref{split}), we subtract from each of the three folded light curves a part of the local average, i.e., $b F_j$. 

We take into account that the plasma temperature of the absorbing wind changes in response to the changing flux within the hard state. Given that the X-ray flux in the hard state of Cyg X-1 is correlated with the 15 GHz flux as $F_{\rm X}\propto F^{1/1.7}$, we adjust accordingly the normalization of $L_{\rm X}$ and $L_{\rm ion}$. 

Then, fitting the model of equation (\ref{tau_ff}) for $b=0$, $i=31\degr$, $f=0.060$, we have obtained the location of the emission at $z_j/a=0.54\pm 0.03$, $0.64\pm 0.03$, $0.72\pm 0.05$, which imply $D_j=0.35\pm 0.02$, $0.26\pm 0.02$ and $0.20\pm 0.02$, and $\bar{\tau}_j=0.32\pm 0.01$, $0.26\pm 0.01$, $0.20\pm 0.01$, for $j=1$, 2, 3, respectively. The folded light curves and the fits are shown in Fig.\ \ref{f:orb_mod15_var}. At $b=0.5$, $f=0.1$, $z_j/a=0.59\pm 0.02$, $0.70\pm 0.02$, $0.79\pm 0.04$, which imply $D_j=0.35\pm 0.01$, $0.26\pm 0.01$ and $0.20\pm 0.01$, and $\bar{\tau}_j=0.72\pm 0.01$, $0.57\pm 0.01$, $0.45\pm 0.01$, for $j=1$, 2, 3, respectively. This $z(F)$ dependence is shown in Fig.\ \ref{f:z_F}. We fit it as a power law, $z\propto F^v$, and obtain $v\simeq 0.51\pm 0.10$, i.e., the statistical significance of an increase of $z$ with $F$ is $5\sigma$. We see that the orbital modulation depth strongly decreases with increasing $F$, also at a high statistical significance. Thus, these calculations confirm the dependence of the radio emission region along the jet on the jet power in relativistic electrons with a high statistical significance. 

As discussed above, we assume that the long term variability occurs in the same way for the modulated and unmodulated parts of the radio flux. We can then write the dependence of equation (\ref{zF}) for the intrinsic flux, equation (\ref{fintr}), and estimate,
\begin{equation}
z(F_{\rm intr})\simeq z_0 \left(F_{\rm intr}\over F_0\right)^q,
\label{z_F}
\end{equation}
where $z_0=z_2$, $F_0=F_2[b+(1-b)\exp(\bar{\tau}_2)]$, and 
\begin{equation}
q\simeq {\ln (z_3/ z_1)\over 
\ln (F_3/ F_1)+\ln\displaystyle{{b+(1-b)\exp(\bar{\tau}_3)\over 
b+(1-b)\exp(\bar{\tau}_1)}}}.
\label{q}
\end{equation}
We obtain $q\simeq 0.62\pm 0.21$, $0.70\pm 0.14$ for ($b=0$, $f=0.06$), ($b=0.5$, $f=0.1$), respectively. In the latter case, we also fit the $z(F_{\rm intr})$ dependence, and obtain $q\simeq 0.72\pm 0.14$, i.e., an almost identical value to that of equation (\ref{q}). These values are somewhat higher than the theoretical prediction of equation (\ref{zF}). This may reflect either the approximate character of the assumed disc model of BK79, the fitted model being too approximate, or both.

We then fit the 8.3 GHz and 2.25 GHz data, see Fig.\ \ref{f:orb_mod}(b--c). The obtained parameters at ($b=0$, $f=0.06$) are $z/a\simeq 1.79\pm 0.22$, $6.17\pm 1.60$ ($\chi^2_\nu\simeq 9.5/9$, 3.9/9), which result in $D\simeq 0.11\pm 0.03$, $0.04\pm 0.03$, and $A=0.77\pm 0.04$, $0.69\pm 0.13$, respectively. The phase lags are $\phi_{\rm min}/2\upi\simeq 0.15\pm 0.05$, $0.22\pm 0.13$, respectively. At ($b=0.5$, $f=0.1$), $z/a\simeq 1.87\pm 0.16$, $6.32\pm 1.16$ ($\chi^2_\nu\simeq 18.7/9$, 7.6/9), which result in $D\simeq 0.11\pm 0.03$, $0.04\pm 0.03$, and $A=0.73\pm 0.04$, $0.64\pm 0.12$, respectively. (At $f=0.2$, $z/a\simeq 2.78\pm 0.21$, $8.84\pm 1.58$, respectively.) The phase lags are unchanged. These data cover a much shorter interval in the hard state than the Ryle/AMI 15 GHz data do, and have been taken with a substantially lower sensitivity. Thus, the resulting folded light curves are much more noisy, especially that for 2.25 GHz, which has the parameters only weakly constrained. Fitting in the log space at 2.25 GHz, 8.3 GHz and 15 GHz with $z\propto \nu^{-n}$, we find $n\simeq 1.4\pm 0.1$, $n\simeq 1.2\pm 0.1$ for $f=0.1$, 0.2, respectively. This is somewhat more than the prediction of the model of BK79, which may be due to either the departures of the jet in Cyg X-1 from that model and/or our fitting model being too approximate.

\section{Effect of free-free absorption on the radio/X-ray correlation}
\label{correlation}

\begin{figure*}
\centerline{\includegraphics[width=11.cm]{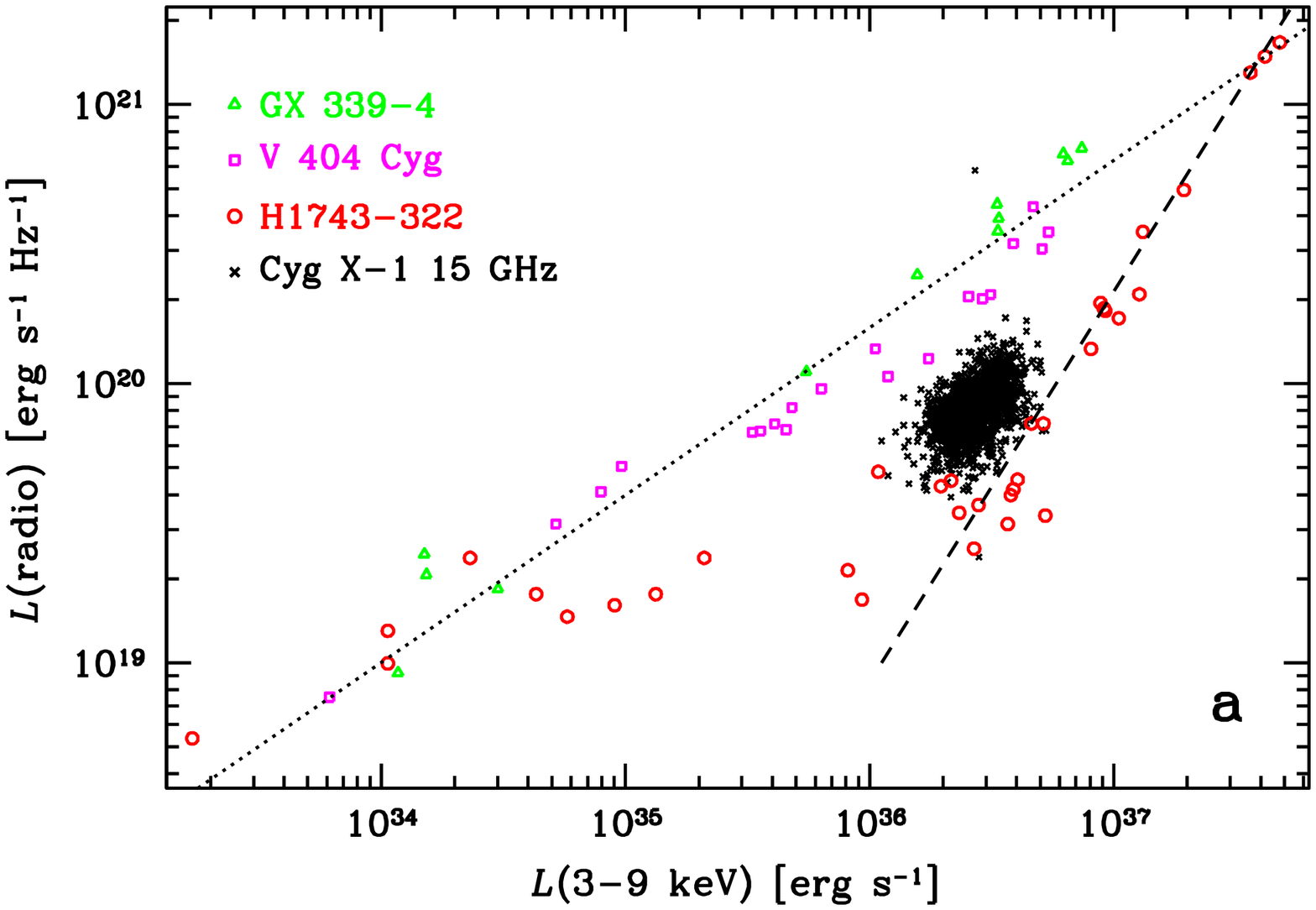}} 
\centerline{ \includegraphics[width=6.11cm]{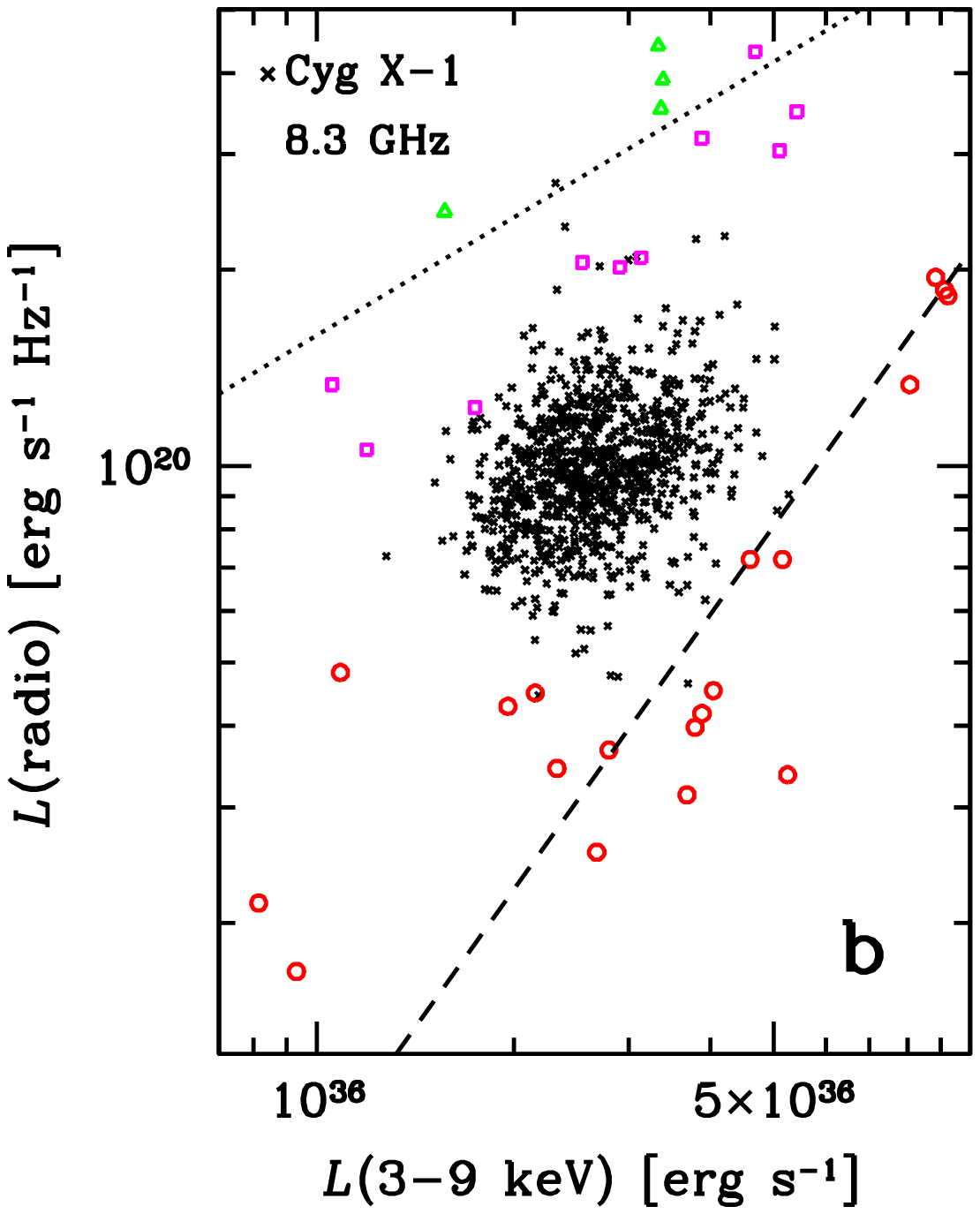}
 \includegraphics[width=4.8cm]{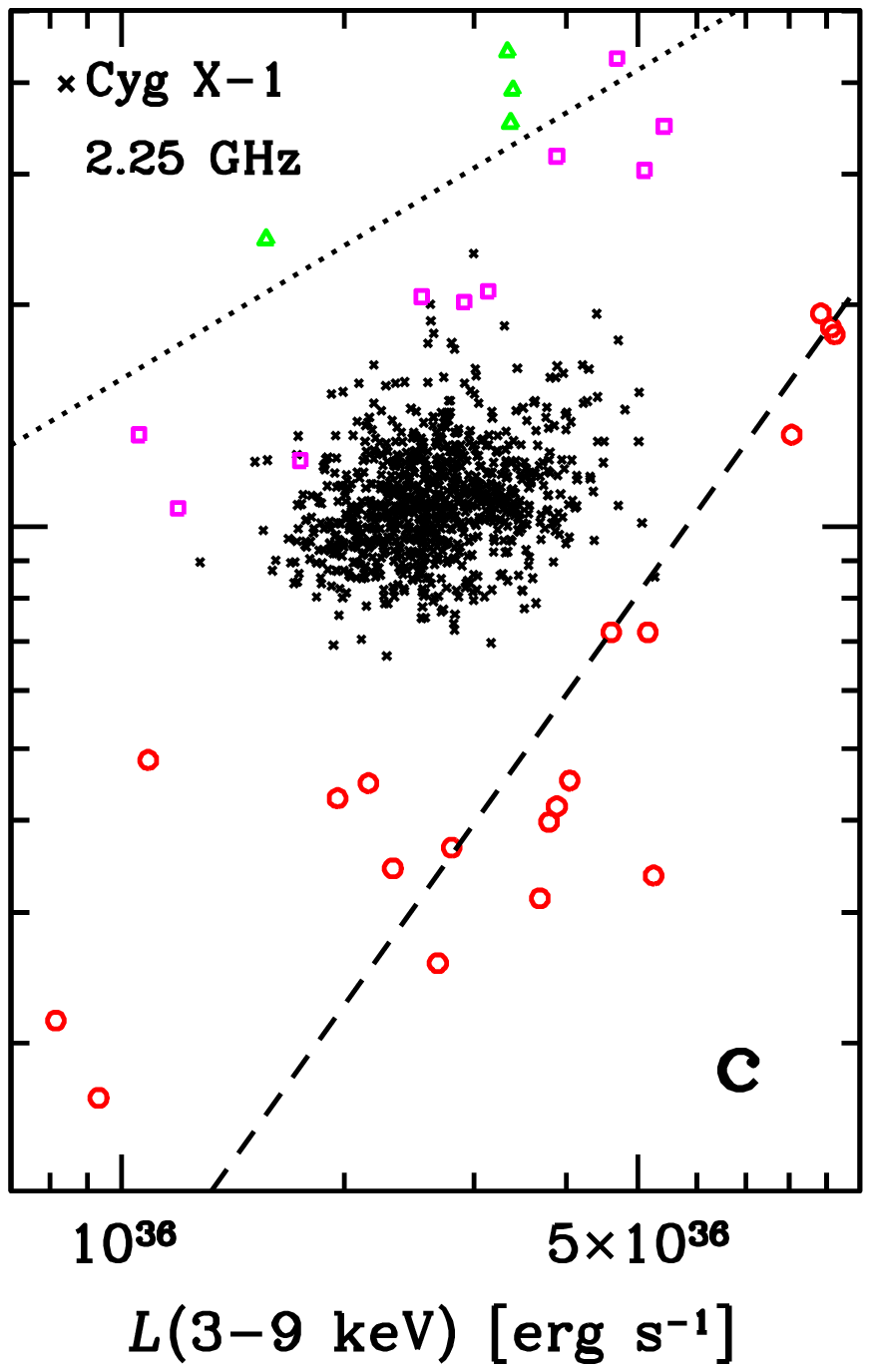}}
\caption{Comparison of the radio/X-ray correlation for Cyg X-1 in the hard state corrected for free-free absorption (assuming $b=0.5$, $f=0.1$, $d=2$ kpc; black crosses) with those for the LMXBs GX 339--4 (\citealt{corbel03}; green crosses), V404 Cyg (\citealt{corbel08}; magenta squares) and H1743--322 (\citealt{coriat11}; red circles). The dotted and dashed lines have $p=0.6$ and $p=1.4$, respectively, approximately fitting the two apparant branches of the correlation for the three sources. See Paper I for the un-corrected Cyg X-1 data. The Cyg X-1 radio data are for (a) 15 GHz, (b) 8.3 GHz, and (c) 2.25 GHz. 
}
\label{f:correlation}
\end{figure*}

Cyg X-1 in all spectral states shows a power-law correlation between the radio and X-ray fluxes, $F \propto F_{\rm X}^p$, where $F_{\rm X}$ is either the flux in an X-ray energy interval, the bolometric flux, $F_{\rm bol}$, or the flux in the Comptonization X-ray component, see Paper I. In particular, the correlation of the 15 GHz flux with $F_{\rm bol}$ in the hard state has the index of $p_{\rm bol}=1.68\pm 0.11$ for the simultaneous ASM and BAT data. Qualitatively, the hard-state correlation in Cyg X-1 is similar to that seen in the hard state of low-mass X-ray binaries (LMXBs) containing black-holes, see, e.g., \citet{gfp03}, \citet{corbel00,corbel03,corbel04,corbel08}, \citet{coriat11}. However, the correlation in Cyg X-1 is significantly steeper than that in LMXBs, e.g., $p_{\rm bol}\simeq 0.8$ in GX 339--4 \citep{z04}. A part or whole of this difference is due to the effect of free-free absorption of the radio emission in the stellar wind. 

An observed correlation index, $p= {\rm d}\ln F/{\rm d}\ln F_{\rm X}$, where $F_{\rm X}$ is either the flux in an X-ray band, the bolometric flux or the flux emitted by hot electrons (see Paper I), is related to its corresponding intrinsic index, $p_{\rm intr}={\rm d}\ln F_{\rm intr}/{\rm d}\ln F_{\rm X}$, by
\begin{equation}
p_{\rm intr}= r p,\quad r= {{\rm d} \ln F_{\rm intr}\over {\rm d}\ln F}
=1+{{\rm d}\ln\left[ b+(1-b)\exp(\bar{\tau})\right] \over {\rm d}\ln F}<1.
\label{index}
\end{equation}
So, the effect of removing free-free absorption of the radio fluxes is to reduce their variability range, without affecting the variability range of the X-ray flux. Note that the factor $r$ is the same regardless of the choice of the energy range for the X-ray flux.

The coefficient $r$ can be determined from the results shown in Figs.\ \ref{f:orb_mod15_var}--\ref{f:z_F}. We estimate it as $r\simeq 1+\Delta \ln\left[ b+(1-b)\exp(\bar{\tau})\right]/\Delta \ln F$. We obtain $r\simeq 0.79\pm 0.02$, $0.71\pm 0.02$ for ($b=0$, $f=0.06$), ($b=0.5$, $f=0.1$), respectively. This implies $p_{\rm intr,bol}\simeq 1.33\pm 0.10$, $1.19\pm 010$, respectively, for the correlation with the bolometric flux. We note that although the wind absorption is significantly stronger for the $b=0.5$ case, it acts only on a 1/2 of the radio flux.

We can also directly calculate the effect of the wind absorption on an individual measurement of the 15 GHz flux, $F$, performed at an orbital phase, $\phi$.  Assuming $f$ and $i$ and using the dependence of $z(F_{\rm intr})$ of equation (\ref{z_F}) allows us to write $F_{\rm intr}$ as,
\begin{equation}
F_{\rm intr}= F\left\{b+(1-b)\exp\left[\tau(z,\phi-\phi_{\rm min})\right]\right\} =F_0\left(z \over z_0\right)^{1/q},
\label{F_intr}
\end{equation}
which we numerically solve for $z$, which yields $F_{\rm intr}$. The resulting $F_{\rm intr}$ points for $b=0.5$, $f=0.1$ are shown together with the 3--9 keV fluxes in Fig.\ \ref{f:correlation}(a). Fig.\ \ref{f:correlation}(a) also compares the radio/X-ray correlation for Cyg X-1 with that in three LMXBs (from \citealt{coriat11}). This figure is similar to one shown in Paper I, which, however, showed the Cyg X-1 radio fluxes as observed. We see that the Cyg X-1 points have now the slope of $p\simeq 1.2$, relatively similar to that of the LMXB H1743--322 \citep{coriat11}. 

The 8.3 GHz and 2.25 GHz data do not allow us to directly make similar estimates, given the poor statistics of the folded light curves, see Fig.\ \ref{f:orb_mod}(b--c). Since the radio spectrum is approximately flat, $F(\nu) \propto \nu^0$ \citep{fender00}, we assume that the $\Delta \ln F_{\rm intr}$ estimated for 15 GHz is accompanied by changes of $z/a$ at the other frequencies proportional to those inferred at 15 GHz, i.e., we use equation (\ref{z_F}) with the value of $q$ obtained for 15 GHz, but $z_0$ and $\bar\tau$ are from the fits to the 2.25 GHz and 8.3 GHz data. Figs.\ \ref{f:correlation}(b--c) show the resulting $F_{\rm intr}$ points for $b=0.5$, $f=0.1$. Given the strong noise in those data, it is difficult to accurately determine the slope formed by those points, but it appears similar to that found for the 15 GHz data. 

However, given the observational uncertainty regarding the actual wind density in the polar regions and possible systematic uncertainties of the fitted model, the case with a higher wind density is also possible, as discussed in Section \ref{cygx1}. For example, for ($b=0.5$, $f=0.2$), $r=0.39\pm 0.06$ at 15 GHz, which corresponds to $p_{\rm intr,bol}\simeq 0.65\pm 0.07$. This would make Cyg X-1 similar to objects in the main branch of the radio/X-ray correlation, e.g., GX 339--4. Thus, our results do not allow us to unambiguously determine the slope of the intrinsic radio/X-ray correlation in Cyg X-1. On the other hand, if we assume that Cyg X-1 in not a unique object, the above constrains the wind density in the polar region to $f\la 0.2$.

\section{Orbital modulation as a function of frequency}
\label{high_nu}

Our results for the 2.25--15 GHz data allow us to make some predictions for the orbital modulation and attenuation by the wind of jet emission at other frequencies, especially higher ones. As estimated in Section \ref{correlation}, $z\simprop \nu^{-1.4}$--$\nu^{-1.2}$. Given the implied low values of $z/a$ for high-frequency radio emission, it may originate within the dense, focused, wind, which would result in much higher wind density than in the polar regions, and hence much higher depths of orbital modulation than those we estimate here. 

\begin{figure}
\centerline{\includegraphics[width=6.8cm]{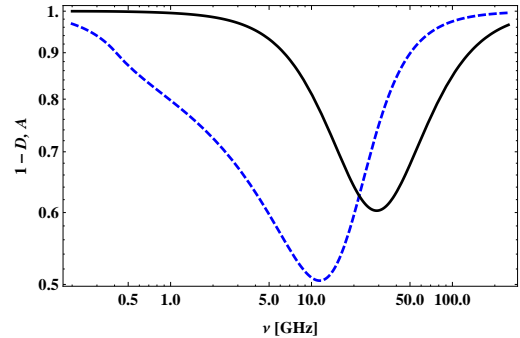}}
\caption{The orbital modulation depth (plotted as the maximum relative attenuation, $1-D$, solid curve) and the average attenuation (dashed curve) as a function of frequency predicted by the model fitted to the 15 GHz data with $b=0.5$, $f=0.2$, $i=31\degr$, and $z\propto \nu^{-1.2}$ based on the 2.25--15 GHz data.
}
\label{f:D_A_vs_nu}
\end{figure}

Still, including the focused wind would introduce additional free parameters to the model, in particular, the photon path going through different wind components. Instead, we assume $f=0.2$, which, as discussed in Sections \ref{cygx1}--\ref{correlation}, approximately corresponds to the highest density of the polar wind compatible with the observational constraints. Our predictions are shown in Fig.\ \ref{f:D_A_vs_nu}. It shows the modulation depth (plotted as $1-D$) and the average attenuation as functions of frequency. We see that the maximum modulation depth occurs for $\nu\simeq 30$ GHz, and it is relatively small at higher frequencies. Also, the attenuation is relatively moderate, $A\ga 0.5$, at all frequencies. 

We see that we predict much lower modulation depths and attenuation than SZ07. The cause for these differences is that our models yield a much lower wind density in the absorbing region than that assumed by SZ07, which is supported by both the current observational results \citep{gies08}, as well as the results of our study of the radio/X-ray correlation in Cyg X-1. Our results may explain the lack of observed modulation at 146 GHz and the 2--220 GHz spectrum being approximately a single power law \citep{fender00}. Still, given the uncertainty related to the effect of the focused wind, it would be of high importance to observationally study the orbital modulation of high-frequency radiation.

\section{Discussion and Conclusions}
\label{conclusions}

We have studied free-free absorption in an optically-thin X-ray irradiated stellar wind. We have taken into account Compton heating and cooling, photoionization heating, bremsstrahlung and line cooling, and advection. We have derived relatively simple formulae for the free-free optical depth, in Section \ref{wind}. We have also considered constraints from eclipses, counter-jets (Appendix \ref{eclipses}) and Compton scattering (Appendix \ref{scattering}).

We have then considered the current constraints on the parameters of the Cyg X-1 binary and its stellar wind (Section \ref{pars}). As found by \citet{gies08}, the wind is strongly anisotropic, with the density close to the orbital plane and in the region shadowed by the donor much higher than that in the polar regions, which are responsible for absorption of the radio emission of the jet. 

In Section \ref{radio}, we have discussed the radio data, including the available information about the extended character of the emission. It appears that a relatively high fraction of the radio emission is emitted far away from the binary orbit, and thus it is not orbitally modulated by the wind. This effect needs to be taken into account in modelling of the modulation.

In Section \ref{cygx1}, we have applied the formalism of Section \ref{wind} to the available radio data from monitoring of Cyg X-1, taking into account the extended character of the emission. In the hard spectral state, the radio light curves show significant modulation at the orbital period, which is due free-free absorption in the wind. We have fitted the folded and averaged light curves at 15 GHz, 8.3 GHz and 2.25 GHz, and obtained the location of the bulk of the emission regions along the jet. In the units of the separation, our best models gave $z/a\simeq 0.6$--1.2, 1.8--2.8, 6.2--8.8, respectively. The height at 15 GHz corresponds to $\sim 3\times 10^{12}\,{\rm cm}\simeq 1.5\times 10^6 R_{\rm g}$, and that at 2.25 GHz is about 10 times higher. These locations, corresponding to the heights at which the jet becomes optically thin to free-free absorption, appear to follow a law relatively close to $z\propto \nu^{-1}$, expected theoretically (BK79). 

On the other hand, \citet{heinz06} in his estimates of the power of the jet in Cyg X-1 based on the model of BK79 used the VLBA images of S01, which imply that 50 per cent of the 8.4 GHz emission comes from distances $\ga 1.5\times  10^{14}$ cm, which is $\sim 50 a$. If the entire jet would follow the emission profile of BK79, only a tiny fraction of the emission would take place at $z\sim a$ (see fig.\ 1 of \citealt{heinz06}), and the resulting orbital modulation would be tiny and probably unobservable, whereas it is at the level $\sim$10 per cent. Given that the distance of the emission should scale as $\nu^{-1}$ (BK79), it would have been even more difficult to explain the $\sim$30 per cent modulation of the 15 GHz emission. A likely solution to this problem is that the observed resolved emission comes from a secondary dissipation event in the jet whereas the emission of the inner jet, at $z\sim a$, is comparable to the resolved one, and it may follow the model of BK79. This would also solve the problem of that jet power calculated by \citet{heinz06} being much lower than that inferred from the observed large-scale optical nebula, presumably powered by the jet \citep{gallo05,russell07}. Furthermore, this would solve the problem of the jet base inferred from the VLBA measurements being an order of magnitude higher than that theoretically expected \citep{rahoui11}.

Recently, \citet{zls12} have calculated the location the jet base in Cyg X-1 (defined by the lowest distance at which significant emission takes place) at $\sim 2\times 10^3 R_{\rm g}$. Using the measurement of the turnover frequency of $\simeq 0.1$ eV of \citet{rahoui11}, and assuming $z\propto \nu^{-1}$ (BK79), gives a location of the 15 GHz emission in agreement with our results.

Our best data, at 15 GHz, allow us to study the orbital modulation as a function of the radio flux. We have found a strong dependence of the modulation depth, changing from $D=0.35\pm 0.01$ to $0.20\pm 0.01$ when the flux increases by a factor of about 2. This is naturally interpreted as the height corresponding to the bulk of the emission increasing with the radio flux as a consequence of the corresponding increase of the jet power. In our case, we find $z$ approximately proportional to the power of about $0.7\pm 0.1$ of the intrinsic radio flux (i.e., that before free-free absorption), which dependence is somewhat stronger than the relation in the BK79 model.

Based on the above results, we have estimated the effect of free-free absorption on the form of the hard-state radio/X-ray correlation in Cyg X-1, which correlation was the subject of Paper I. We find the absorption has a substantial effect, but, given the uncertainty about the wind density in the polar regions of Cyg X-1, we cannot accurately quantify this effect. This effect changes the correlation index, $p$ (defined as $F\propto F_{\rm X}^p$), from about 1.7, found in Paper I, to about 0.7--1.3. Thus, Cyg X-1 may have a similar correlation to that of the LMXB H1743--322, which has $p\sim 1.4$ \citep{coriat11}, or a number of other LMXBs, e.g., GX 339--4 \citep{corbel03} and V404 Cyg \citep{corbel08}, where the correlation index is $p\sim 0.6$--0.8.

We also discuss predictions for the orbital modulation and attenuation by free-free absorption of radio emission at frequencies higher than 15 GHz. Given the relatively low density of the polar wind in Cyg X-1, these effects are likely to be much weaker than estimated before (in SZ07), which may explain the apparent lack of orbital modulation at $\nu= 146$ GHz, and the 2--220 GHz radio spectrum of Cyg X-1 being represented by a single power law \citep{fender00}. Still, it would be highly desirable to actually measure the modulation depth at high frequency, which would give us significant constraints on the structure of both the wind and the jet in Cyg X-1. In particular, this would test the possible location of the high-frequency radio emission within the focused wind component. 

Finally, we mention that another high-mass X-ray binary with strong wind, Cyg X-3 (which is likely to contain a black-hole) does not show any orbital modulation of its very strong radio emission. As discussed by SZ07, this may be due to the binary orbit of that system being $\sim$10 times smaller than that in Cyg X-1. If the jet of Cyg X-3 is similar in size to that of Cyg X-1, this would imply values of $z/a$ $\sim$10 times higher, which is likely to reduce the depth of the radio modulation to unobservable values.

\section*{ACKNOWLEDGMENTS}

Thanks are due to F. Yuan for very valuable suggestions. I also thank J. Miko{\l}ajewska and J. Zi{\'o}{\l}kowski for valuable discussions, and the referee for valuable comments. This research has been supported in part by the Polish NCN grants N N203 581240, N N203 404939, and 362/1/N-INTEGRAL/2008/09/0. The AMI Arrays are operated by the University of Cambridge and supported by the STFC. The Green Bank Interferometer is a facility of the National Science Foundation operated by the NRAO in support of NASA High Energy Astrophysics programs.

\appendix
\section{Eclipses and the counter-jet}
\label{eclipses}

The jet emission can be obscured by the donor provided $z<R_*$. The lack of such eclipses implies,
\begin{equation}
i<\upi/2-\arctan {a\over z}+\arccos{R_*\over (a^2+z^2)^{1/2}},\quad z<R_*.
\label{ecl}
\end{equation}
In case of the lack of X-ray eclipses ($z=0$), $i< \arccos(R_*/a)$, which implies no eclipses of the jet emission. Thus, equation (\ref{ecl}) is of importance only in systems where X-ray eclipses are observed (e.g., M33 X-7, \citealt{orosz07}), in which case observations of eclipses of the radio emission or their lack would impose a constraint on the location of the radio-emitting region. In the case of Cyg X-1, the lack of X-ray eclipses implies $i< 64\degr$ for $a/R_*\simeq 2.3$. 

There will be no eclipse of the counter-jet emission by the donor for either,
\begin{eqnarray}
\lefteqn{i>\arctan {a\over |z|}+\arcsin{R_*\over (a^2+z^2)^{1/2}},\quad |z|>R_*,}\label{icju}\\
\lefteqn{i<\arctan {a\over |z|}-\arcsin{R_*\over (a^2+z^2)^{1/2}}.\label{icjl}}
\end{eqnarray}
For example, for $a=2.3 R_*$ and $z=-a$, no counter-jet eclipses correspond to $i< 28\degr$ or $i> 63\degr$. In addition, an accretion disc with the outer radius, $r_{\rm o}$, will block the jet emission for $i<\arctan(r_{\rm o}/z)$.

Equation (\ref{tau_ff}) also describes the attenuation of the counter-jet emission (in the constant velocity, constant temperature approximation), with the substitution of $z\rightarrow -z$. Note that then there is a singularity in equation (\ref{tau_ff}) is at $\phi=0$ and $z/a=-\cot i$, which is a true singularity, caused by photon paths going through the centre of the donor. 

Note that the counter-jet emission is both relativistically beamed away from the line of sight and more absorbed than the direct jet emission. Also, the counter-jet emission can be obscured by the star at some combinations of $z$ and $i$, see equation (\ref{icju}). All these factors can account for the lack of detection of the radio counter-jet in Cyg X-1 (S01). 

\section{Compton scattering and bound-free absorption}
\label{scattering}

We now consider scattering or absorption proportional to the density (which is the case for bound-free absorption) of photons emitted by the jet and accretion flow. In the latter case, we require that the medium temperature and ionization parameter are constant along the photon length, which is a commonly adopted assumption in calculating bound-free absorption.

The optical depth is then,
\begin{equation}
\tau= \tau_0 \int_0^\infty \left(r\over a\right)^{-2} \left(1-{R_* \over r}\right)^{-\beta} {\rm d}(l/a),
\label{tau_sc_v}
\end{equation}
where, in the case of Compton scattering, $\tau_0=\sigma_{\rm C} n_0 a \mu_{\rm i}/\mu_{\rm e}$, and $\sigma_{\rm C}$ is the scattering cross section. We can integrate equation (\ref{tau_sc_v}) numerically, or, assuming $v=v_\infty$, we obtain,
\begin{equation}
{\tau\over \tau_0} = {\upi/2+ \arctan(s u^{-1/2})\over u^{1/2} }.
\label{tau_sc}
\end{equation}
As before, $u=0$ for $\phi=\upi$, $z/a=\cot i$, and equation (\ref{tau_sc}) appears to have a singularity. However, $\tau$ is finite, $\tau/\tau_0 = \sin i$. The other singularity at $\phi=0$, $z/a=-\cot i$, is a true one. At $z=0$, equation (\ref{tau_sc}) yields \begin{equation}
{\Delta \tau\over \tau_0}={2i\over \cos i}.
\label{taut}
\end{equation}

Cyg X-1 shows an orbital modulation at hard X-rays, at which energies the bound-free absorption is negligible, and which is most likely due to Compton scattering of photons away from the line of sight by the wind electrons. The source of X-rays in this case is close to the compact object, and thus we may set $z=0$. However, the assumption of $v=v_\infty$ leads to a significant underestimate of the actual $\tau$. Thus, we use the numerical integration of equation (\ref{tau_sc_v}). For the parameters of Cyg X-1 listed above, we obtain $\tau_0 \simeq 8.4\times 10^{-3} f$, where the Thomson cross section has been used. The results of \citet{l06} and \citet{zps11} yield the orbital modulation depth at $\sim$20--100 keV hard X-rays (which photon interactions are dominated by the Thomson scattering) of $\simeq 0.03$--0.04. We find $D\simeq 0.02f$ at $i=31\degr$, which implies $f\simeq 1.5$--2. This appears to account for the wind density being enhanced near the orbital plane (see Section \ref{pars}).

\label{lastpage}

\end{document}